\documentclass[useAMS,usenatbib,usegraphicx]{mn2e}

\usepackage{amsmath} 

\title[Mkn 1 ionized cloud]{An [O III] search for extended emission around AGN with H I mapping: a distant cloud ionized by Mkn 1}
\author[E.D Knese et al.]
{Erin Darnell Knese$^{1,2,3}$\thanks{E-mail: erin.knese@huschblackwell.com},
William C. Keel$^{1,3}$ \thanks{E-mail: wkeel@ua.edu},  Greg Knese$^{4}$,
Vardha N. Bennert$^{5}$, 
\newauthor Alexei Moiseev$^{6}$, Aleksandra Grokhovskaya$^{6}$, and Sergei N. Dodonov$^{6}$\\
$^{1}$Department of Physics and Astronomy, University of Alabama, Box 870324, Tuscaloosa, AL 35487, USA\\
$^{2}$Current address: Husch Blackwell, 190 Carondelet Plaza, Suite 600, St. Louis, MO 63105, USA\\
$^{3}$SARA Observatory\\
$^{4}$Department of Mathematics, Washington University in St. Louis, One Brookings Drive, St. Louis, MO 63130, USA\\
$^{5}$Department of Physics, California Polytechnic State University,  San Luis Obispo, CA 93407, USA\\
$^{6}$Special Astrophysical Observatory, Russian Academy of Sciences, Nizhny Arkhyz 369167, Russia\\
}

\begin{document}

\date{}

\pagerange{\pageref{firstpage}--\pageref{lastpage}} \pubyear{2019}

\maketitle

\label{firstpage}

\begin{abstract}
Motivated by the discovery of large-scale ionized clouds around AGN host galaxies, and particularly the large fraction of those which are consistent with photoionized gaseous tidal debris, we have searched for [O III] emission over wide fields around a set of Seyfert galaxies previously mapped in H I, many of which show extended gas features. The detection threshold was designed to reach mean emission-line surface brightness 10 times fainter than seen in such AGN-ionized clouds as Hanny's Voorwerp, so that similar structures at larger distances (and ages) could be detected. Of 24 Seyfert galaxies, we find one extended emission feature, a discrete cloud projected $\approx 12$ kpc from the center of Mkn 1 and
spanning a transverse extent of 8 kpc. In addition, we identify several potential ``emission-line dots" (ELdots), compact objects which may be outlying,
relatively isolated star-forming regions. Optical spectroscopy of the Mkn 1 cloud confirms its redshift association with the Mkn 1 - NGC 451 
galaxy pair, 
shows it to closely match the kinematics of nearby H I, and reveals emission-line ratios requiring photoionization by the AGN at roughly
the direct observed luminosity of the nucleus.. 
Given the small fraction of H I features with detected [O III] emission, we constrain the typical opening angle of ionization cones in Seyfert galaxies 
to be of order $20^\circ$, on the assumption that active episodes are long compared to the light-travel times involved. An appendix presents a derivation of an analytical expression for the probability of intersection of a cone with randomly oriented arcs, approximating the geometry of H I clouds and tails exposed to ionization cones. For the entire sample, the full opening angle of bicones must be
$<20^\circ$ if the AGN are continuously bright for scales longer than the light-travel times
across the H I structures. Since many ionization cones are observed to be much broader than this, our low
detection fraction may add to evidence for the ubiquity of strong variations in AGN luminosity on scales
$10^{4} - 10^5$ years.

\end{abstract}


\section{Introduction}

A subset of active galactic nuclei (AGN) has been known to be accompanied by extended emission-line regions (EELRs) 
since pioneering discoveries in the 1980s. As reviewed by, for example, \cite{Wilson} and \cite{Stockton}, these
often take the form of ionization cones, and in some cases extend tens of kpc outside the AGN host galaxy itself.
EELRs allow us to trace the pattern of emerging radiation, characterize AGN which are strongly
obscured along our line of sight, or have undergone dramatic luminosity changes over millennia. These applications
have seen new use with recent discoveries of EELRs around AGN which appear too faint to account for their ionization
level, requiring either strong obscuration or strong variability (over the light-travel times between nucleus and EELR) 
to explain this mismatch. In particular, Hanny's Voorwerp
(\citealt{Lintott2009}, \citealt{HVHST}, \citealt{Kevin2010}) is a galaxy-scale highly-ionized cloud for which the nearby
AGN in IC 2497 fails to be able to account for its ionization by factors 20-100 even when its spectral energy distribution up to hard X-rays 
is modeled \citep{Sartori}. Following the discovery of this object by Galaxy Zoo participant Hanny van Arkel, a dedicated search
by additional Galaxy Zoo participants found another 19 AGN with similar ionized clouds projected $>10$ kpc from the 
nuclei \citep{mnras2012}, of which 8 have a substantial energy deficit from the AGN as observed \citep{Keel2017b}. Additional
studies have identified analogous objects at both lower and higher AGN luminosity (\citealt{greenbeans}, \citealt{ngc7252}).

These findings have allowed the beginnings of a picture of AGN changes over time spans $10^4-10^5$ years,
which can connect to longer time spans inferred from simulations and statistics of AGN in interacting hosts,
and with the shorter timescales known from direct observation (reverberation studies and the growing number of ``changing-look" 
AGN). If the AGN surrounded by EELRs are typical, their radiative output is characterized by luminous phases
$\approx 10^5$ years long with low-luminosity interludes, possibly associated with transitions between the accretion
output being mostly radiative or mostly kinetic (\citealt{mnras2012}, \citealt{Keel2017b}, \citealt{Kevin2015}). The
``changing-look" AGN (e.g., \citealt{LaMassa}, \citealt{Runnoe}, \citealt{Ruan}), which change their nuclear luminosity significantly within only a few years, show that such rapid changes in the
AGN luminosity occur. Changes traced by light-travel time to clouds tens of kpc from the AGN, in contrast,
must represent millennium-scale durations when the average luminosity stays high or low.

Since almost all EELRs have been found around galaxies with extant (if sometimes low-luminosity) AGN, and
surrounding gas is a prerequisite to their occurrence, we have carried out a series of imaging surveys in the strongest optical emission 
line from EELRs, [O III] $\lambda 5007$. The sample described here consists of Seyfert galaxies mapped in H I by \cite{Kuo}, so
we know which ones have extended H I discs or tails, and roughly what fraction of each AGN host is surrounded
by such gas. This lets us use the detection fraction to estimate the covering factor of escaping gas
(opening angle of ionization cones) combined with the fraction of the time the typical AGN is in a luminous state over 
the scales spanned by light-travel times to the outermost gas. We also consider in detail the properties of the sole EELR
discovered, near Mkn 1.

In quoting luminosities and
sizes, we adopt $\rm H_0= 72$ km s$^{-1}$ Mpc$^{-1}$ and flat cosmological geometry.

\section{Sample selection and observations}

\subsection{Sample construction}

Our sample consists of the Seyfert galaxies observed by \cite{Kuo} in their study of H I structures around active
galaxies, including data on the four objects presented earlier by \cite{Greene}. 
We also observed the H II-region nuclei Mkn 1158 and 1510 included in \cite{Kuo}, but do not analyze their fields here.
This leaves 26 Seyfert galaxies in our sample (Table \ref{tbl-1}). They lie at redshifts $z=0.015-0.020$; for guidance, we include Hubble types from
\cite{Kuo} and AGN classifications from NED\footnote{The NASA Extragalactic Database, http://ned.ipac.caltech.edu.}. 
As our further work has progressed, this is 
now the first phase of a multipart program we call TELPERION\footnote{The name is mostly intended for readers of Tolkien's work
on the First Age of Middle-Earth, to connote long-vanished brilliance, but also stands for Tracing Emission Lines to
Probe Extended Regions Ionized by Once-active Nuclei}. Further phases now in progress encompass merging galaxies in the Toomre sequence,
luminosity-selected AGN, luminous galaxies independent of AGN, and a larger merging sample from the Galaxy Zoo \citep{Lintott2008} analysis 
by \cite{Darg}.

\begin{table*}
\begin{center}
\caption{Properties of Sample Galaxies}  \label{tbl-1} 
\begin{tabular}{lccccccc}
\hline
Name & z & M$_{B}$ & AGN type & Hubble Type & Inclination & F$_{NUC}$(5007) \\
   &   &   & Class & Type & ( $\degr$ ) & (erg cm$^{-2}$ s$^{-1}$) \\
\hline
Mkn 341  & 0.0153 & -21.9 & S & S0-a & 70.5 & $8.3 \times 10^{-14}$ & \\
NGC 266  & 0.0155 & -20.6 & S3b & Sab & 14.5 & $3.3  \times 10^{-14}$ & \\
Mkn 352  &  0.0149 & -19.5 & S1.0 & S0 & 33.9 & $2.5  \times 10^{-15}$ & \\
Mkn 1  &  0.0159 & -19.0 & S2 & Sb & 58.7 & $2.3  \times 10^{-13}$ & \\
NGC 513  &  0.0195 & -21.5 & S1h & Sc & 61.9 & $6.3  \times 10^{-14}$ & \\
Mkn 993  &   0.0155 & -20.0 & S1.5 & Sa & 90.0 & $1.9  \times 10^{-14}$ & \\
Mkn 359  & 0.0174 & -20.2 & NLS1 & Sb & 39.1 & $4.9  \times 10^{-14}$ & \\ 
Mkn 1157  &  0.0152 & -20.1 & S1h & S0-a & 40.2 & $6.1 \times 10^{-14}$ & \\
Mkn 573  &  0.0172 & -20.1 & S1h & S0-a & 28.0 & $4.1  \times 10^{-13}$ & \\
UGC 1395  &  0.0174 & -20.5 & S1.9 & Sb & 54.8 & $2.5  \times 10^{-13}$ & \\
NGC 841  &  0.0151 & -21.4 & S3b & Sab & 64.8 & $3.7  \times 10^{-14}$ & \\ 
Mkn 1040  &  0.0167 & -19.2 & S1.0 & Sbc & 81.1 & $2.5  \times 10^{-14}$ & \\ 
NGC 1167  &   0.0165 & -21.1 & Sy 2 & S0 & 49.0 & $4.3  \times 10^{-17}$ & \\
UCG 3157  &   0.0154 & -19.9 & S2 & Sbc & 35.2 & $6.4  \times 10^{-15}$ & \\   
MS 04595+0327  &   0.0160 & -19.5 & S1 & E & 50.2 & $1.4  \times 10^{-13}$ & \\   
IRAS 05078+1626 &   0.0179 & -19.4 & S1.5 & E? & 39.7 & $1.0  \times 10^{-13}$ & \\
UGC 3995  &  0.0158 & -21.5 & S2 & Sb & 67.6 & $1.1  \times 10^{-14}$ &\\  
Mkn 1419  &   0.0165 & -19.8 & S3 & Sa & 41.9 & $5.3  \times 10^{-15}$ & \\
Mkn 461 &   0.0162 & -20.4 & S2 & Sab & 43.6 & $3.3  \times 10^{-14}$ & \\
IRAS 14082+1347  & 0.0162 & -19.5 & S3 & S? & 53.5 & $3.1  \times 10^{-15}$ & \\   
NGC 5548  &   0.0172 & -20.7 & S1.5 & S0-a & 41.0 & $4.9  \times 10^{-14}$ & \\    
ARK 539  &  0.0169 & -19.9 & S2 & S? & 31.9 & $2.0  \times 10^{-14}$ & \\    
NGC 7469  &   0.0163 & -21.6 & S1.5 & Sa & 30.2 & $5.6  \times 10^{-13}$ & \\    
NGC 7591 &  0.0171 & -21.6 & S & SBb & 67.1 & $6.6  \times 10^{-14}$ & \\    
NGC 7679 &  0.0171 & -21.4 & S2 & S0-a & 59.1 & $5.1 \times 10^{-13}$ & \\ 
NGC 7682 &  0.0171 & -20.7 & S1h & Sab & 22.7 & $2.8  \times 10^{-13}$ & \\ 
\hline
\end{tabular}
\end{center}
\end{table*}

\subsection{Observations: narrowband imaging}

The survey observations were carried out between October 2010 and July 2012 using the remotely-operated
telescopes of the SARA Observatory \citep{SARA}. The narrowband images, and most of the broadband continuum
imaging, used the 1m instrument on Kitt Peak, Arizona (SARA-KP). During this period it was equipped with a $2048\times 2048$-pixel E2V CCD in
an Apogee U42 camera. The pixel scale was 0\farcs 382  pixel$^{-1}$, giving a field 13.04\arcmin\  square. For a centered
galaxy, this meant the images covered (inscribed) projected radii 118-153 kpc over the redshift range of our sample. The
$V$-band continuum images for five galaxies (MS 04595+0327, UGC 1395, NGC 7591, NGC 7679 and NGC 7682)
were obtained using the SARA 0.6m telescope on Cerro Tololo, Chile. The Apogee Alta E6 camera in use then gave a field
10.34 \arcmin\  square at 0\farcs 606 pixel$^{-1}$, covering radial regions out to $\approx100$ kpc from each targeted galaxy.

A narrowband filter centered at 5100 \AA \, with half-transmission points at 5047 \AA \, and 5132 \AA \, captured [O III] 
emission at the redshifts of sample members. The filter is circular with a 50mm diameter, and was fabricated by Custom Scientific\footnote{http://customscientific.com}. Corners of the CCD images are slightly vignetted, an effect well corrected using twilight-sky flat fields. A standard $V$ passband, which has its center of transmission close to the narrowband filter's peak transmission, was used for comparison images in the broadband continuum. 

In the converging f/8 beam of this telescope, we expect the peak response of the [O III] to be shifted blueward by about 10 \AA, which we correct in deriving emission-line fluxes. Our exposures were designed to reach mean emission-line surface brightness at least 10 times fainter than Hanny's Voorwerp,
allowing detection of similar objects farther from the ionizing AGN, as well as objects ionized by less powerful AGN. Exposure sequences were stacked for a 5400 s total exposure in [O III] and an 1800 s total $V$ exposure. Calibration frames were observed at the beginning of each run and standard reduction procedures were carried out using NOAO's $ccdred$ package in IRAF\footnote{IRAF is distributed by the National Optical Astronomy Observatory, which is operated by the Association of Universities for Research in Astronomy (AURA) under a cooperative agreement with the National Science Foundation.} \citep{Tody1986}. This CCD suffered from residual bulk image (e.g. \citealt{cri11}), producing image persistence from bright objects which decayed with an $e$-folding time typically 40 minutes. This effect was mitigated by taking a shorter dark exposure before each new object, subtracting a smoothed and scaled version of this, as well as offsetting the telescope between 
the 3 individual 30-minute exposures with the narrowband
filter. Stacking these multiple exposures also substantially rejected cosmic-ray events and residual 
flat-field imperfections. The individual narrowband exposures were sky-noise limited, so
breaking the total observation into 3 exposures imposed little penalty in signal-to-noise ratio (SNR).

Continuum-subtracted (emission-line) images were produced by scaling and subtracting the combined continuum image from the combined narrow-band image. Scaling factors were determined from flux ratios of stars with known color. 

To detect spatially extended emission regions of low surface brightness, we applied several smoothing algorithms to the emission-line images. Median filtering (box size $\sim 2 \times 2$\arcsec \ ) and Gaussian smoothing ($\sigma\sim3\farcs 5$) over the images improved the detectability of 
structures  in these scales. Detection limits in [O III] emission for structures much larger than these
smoothing windows are typically $2 \times 10^{-17}$ erg cm$^{-2}$ s$^{-1}$ arcsec$^{-2}$.

To estimate the [O III] flux (or surface brightness) of detected objects, we establish a calibration using \cite{lan92} standard stars observed on photometric nights, along with count-rate ratios of stars between broad 
and narrow filters. Following \citet{fsi95}, we converted magnitude zero points  to total flux corresponding to one count/second 
within both broad- and narrow-band filters. The ratio of stellar count rates between filters, and the filters' effective widths, let us determine the flux 
represent by on ADU per unit time in redshifted $\lambda 5007$ emission. This calibration factor, and the fidelity of continuum subtraction,  
depend on the color of any associated
 continuum; extreme colors work less well since we used only a single continuum filter, centered slightly to the red of the narrowband filter. The emission-line flux is corrected for the wavelength-dependent filter transmission, using a correction factor equal to the peak transmission divided by the transmission (at that $z$) of the 5100-\AA\  filter. [O III] fluxes for the AGN were measured from the emission-line images using 
 {\it imexam} from the IRAF {\it images} package, 
within a 5\arcsec aperture with automatic centering turned on. Three iterations were allowed to adjust the fitting radius. Flux counts for all other detected compact emission-line sources were calculated with automatic centering turned off and no allowed adjustments to the fitting radius,
while for diffuse sources we used larger box apertures as appropriate.

To improve our knowledge of the diffuse cloud found in our survey data,
we obtained new images of Mkn 1 with SARA-KP in late 2012 and late 2016, after installation of a new imager using a nominally identical chip, operating at much lower temperature. The new camera from ARC{\footnote{http://www.astro-cam.com} essentially eliminated the thermal noise which had been important in our earlier narrowband images, and eliminated the bulk-charge afterimages. 
A 3.5-hour exposure 
stack in the same $\lambda 5100$ filter goes significantly deeper, revealing structure in the cloud (Fig. \ref{fig-Mkn1SARA}). The image does show residuals 
from removal of the scattered light produced by the 6th-magnitude star HD 7578 just outside the field to the NE, which were
largely reduced along with charge bleeding from other field stars by subtracting a version of the image median-filtered
using a $1 \times 275$-pixel box parallel to the detector $y$-axis. 
Similar data were obtained for NGC 7591 and UGC 3995, allowing us to reject marginal candidate 
detections in those field from the original survey images.

Coordinate mapping and alignment of the images, including a small rotation of the camera between sessions, used 
astrometric solutions with the astrometry.net web interface \citep{Lang}. The two components of the [O III] cloud are centered at 2000 coordinates
$\alpha=$01 16 05.32, $\delta=$ +33$^\circ$ 04\arcmin 50\farcs 0 and
and $\alpha=$01 16 04.79, $\delta=$ +33$^\circ$ 04\arcmin 53\farcs 4.

The estimated total flux in [O III] $\lambda 5007$ is $9.3 \times 10^{-15}$ erg cm$^{-2}$ s$^{-1}$ arcsec$^{-2}$, with a characteristic
surface brightness within the detected area of $\approx 6.6 \times 10^{-17}$ erg cm$^{-2}$ s$^{-1}$ arcsec$^{-2}$.

\begin{figure*} 
\includegraphics[width=140.mm,angle=90]{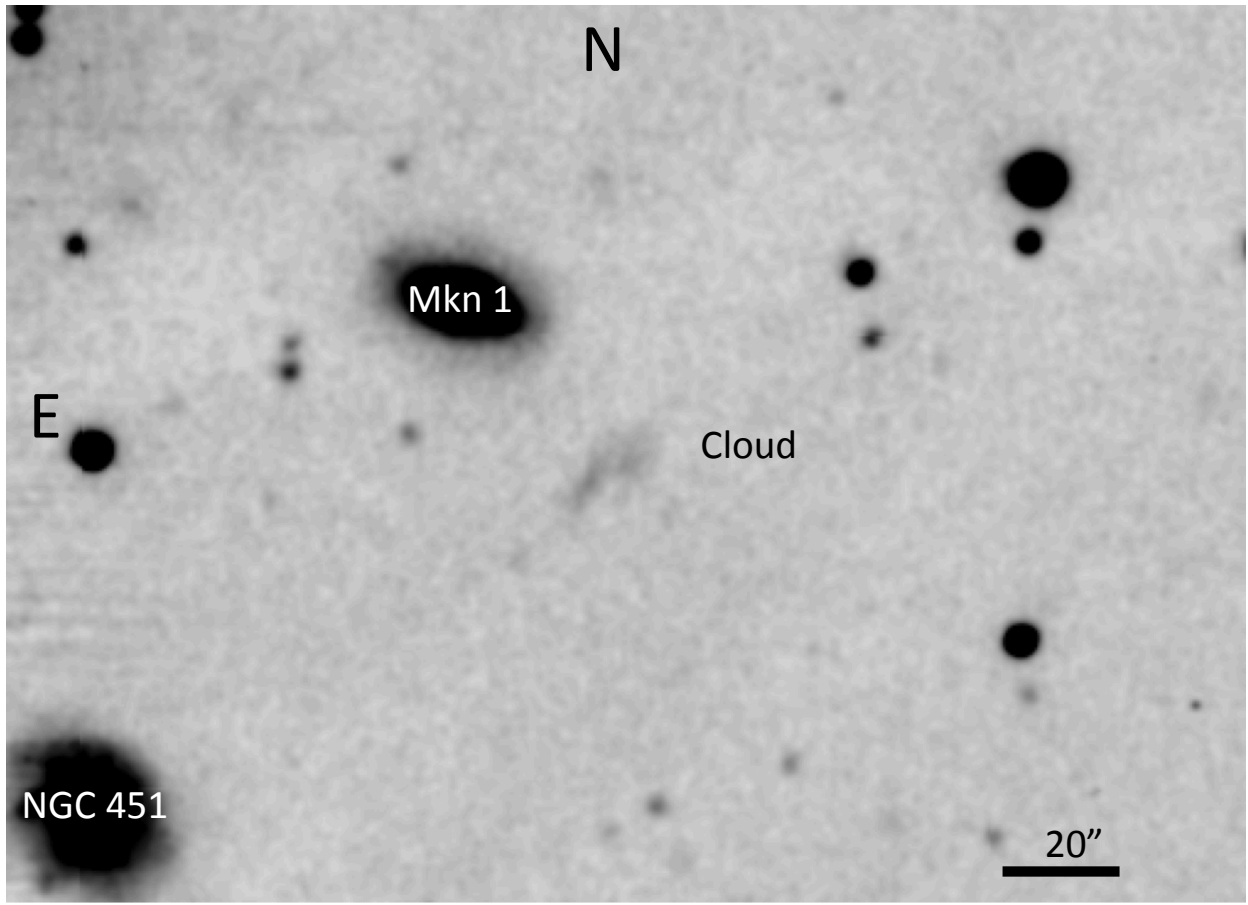} 
\caption{Stacked SARA image of the Mkn 1 field in the 510-nm narrowband filter including 
redshifted [O III] emission. The newly detected ionized cloud appears near the center. This 
image has been smoothed with a Gaussian of 1\farcs 6 FWHM for display.
The region shown is $154 \times 217$ \arcsec with north at the top. Some faint large-scale features result from
imperfect removal of reflections caused by the 6th-magnitude star HD 7578 just off the northeast edge of this image section.}
 \label{fig-Mkn1SARA}
\end{figure*} 

Supplementary imaging of the Mkn 1 field was obtained in redshifted H$\alpha$ (including the adjacent [N II] and [S II] lines), via a filter 
centered at 675 nm with FWHM 25 nm (6000-s exposure), and the $g,r$
continuum bands (3000 seconds each), using the recently commissioned CCD imager on the 1-meter Schmidt telescope of the 
Byurakan Astrophysical Observatory \citep{Dodonov2017}. After trials involving both $g$ and $r$ for continuum subtraction, use of the $r$ band alone 
was found to give the
smallest residuals near Mkn 1 and NGC 451. The subtraction has the main advantage, even for pure emission clouds, of dramatically reducing
the scattered light from the bright foreground star. This image gives evidence of a fainter cloud opposite the SW one we examined
spectroscopically (Fig. \ref{fig-Mkn1Byu}). We do not yet have additional confirmation of its existence; structure in the extended PSF of the star includes
pieces of tangent arcs which would share its orientation.

\begin{figure*} 
\includegraphics[width=134.mm,angle=90]{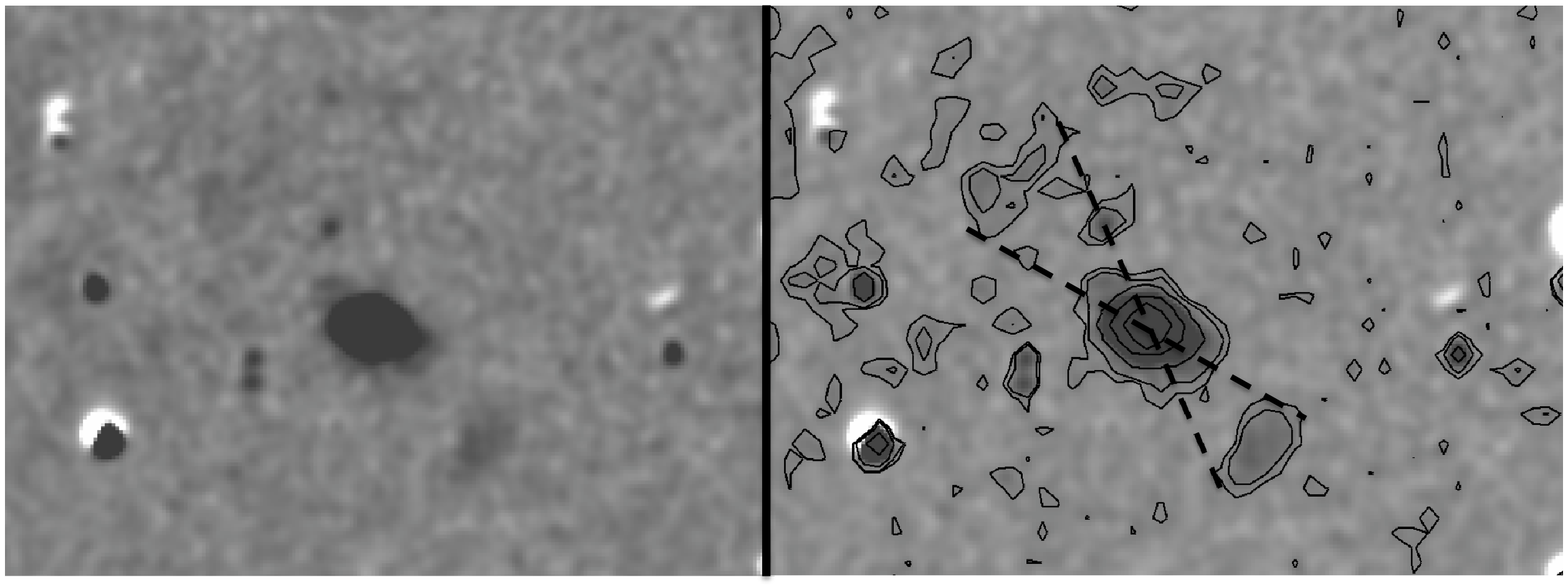} 
\caption{Byurakan H$\alpha$+[N II]+[S II] image of the field of Mkn 1, using $r$ for approximate continuum subtraction and smoothed
with a Gaussian of FWHM 2.5\arcsec to emphasize faint structures. The right-hand panel overlays sparse contours at
logarithmic spacing to show the extent of the emission clouds, with dashed lines connecting them in notional ionization
cones to illustrate the symmetric locations of the confirmed SW and unconfirmed NE clouds. This field spans $132 \times 186$\arcsec
with north at the top.}
 \label{fig-Mkn1Byu}
\end{figure*} 

\subsection {Lick spectroscopy}

To investigate the ionization source of the cloud detected near Mkn 1, we obtained a long-slit spectrum using the
Kast double spectrograph \citep{Kast} at the 3-m Shane telescope of Lick Observatory, on 14 January 2013. The slit was 
oriented
along position angle $304^\circ$, so that when the companion galaxy NGC 451 was centered near one end of the 2\arcsec -wide
slit, the cloud would be near the other end and sampled nearly along its longest axis. Four 30-minute exposures were combined. 
The red channel covered the wavelength range 4640--7400 \AA\  at $\approx 6$ \AA\ resolution, while the blue channel spanned 
3350--4580 \AA\ at $\approx 2$ \AA\ resolution. Subtraction of night-sky
emission lines was compromised by having objects of interest at each end of the slit, a particular problem for H$\beta$.

\subsection{BTA spectroscopy}

We obtained a deeper long-slit spectrum covering the companion galaxy to Mkn 1 (NGC 451) and the ionized cloud,
using the SCORPIO multimode instrument \citep{SCORPIO} at the 6-meter telescope (BTA) of the Russian Academy of Sciences.
A 2-hour exposure on 17 August 2017 was set with the slit at PA $304^\circ$, so acquisition of the companion nucleus
would also put the slit nearly along the major axis of the cloud.
The slit width was 1\farcs 0, and the spectral coverage including the [O II] $\lambda 3727$ and [S II]
$\lambda 6717$ features. The redder $\lambda 6731$ feature of [S II] is compromised by overlap with the 
atmospheric B band at this redshift, and a night-sky emission line. We corrected the data for B-band absorption using the spectrum of
a hot star along the slit, and explored several fitting routines to correct for residual night-sky variations along the slit. The uncertainty
range on the $\lambda 6717/ \lambda 6731$ ratio is still $>1.40$, allowing only values close to the low-density limit.
We adopt a $2 \sigma$ limit $n_e < 26$ cm$^{-3}$, using the {\it temden} task from \cite{ShawDufour} and
assumed temperature $10^4$ K.
 The [O II] $\lambda 3727$ doublet is in a region where the system sensitivity is low, giving a signal-to-noise ratio
too low for an independent density estimate. The spectral scale was
2 \AA\  pixel$^{-1}$, strongly oversampling the spectral resolution of 12 \AA\ FWHM, with the binned pixels spanning 0\farcs 357 along the slit. 
The signal-to-noise ratio of these
data is sufficient to measure He II emission and (marginally) detect [O III] $\lambda 4363$.

Fig. \ref{fig-BTAspectrum} shows the cloud spectrum summed along 65 spatial pixels (23\arcsec\ ).

\begin{figure*} 
\includegraphics[width=120.mm,angle=90]{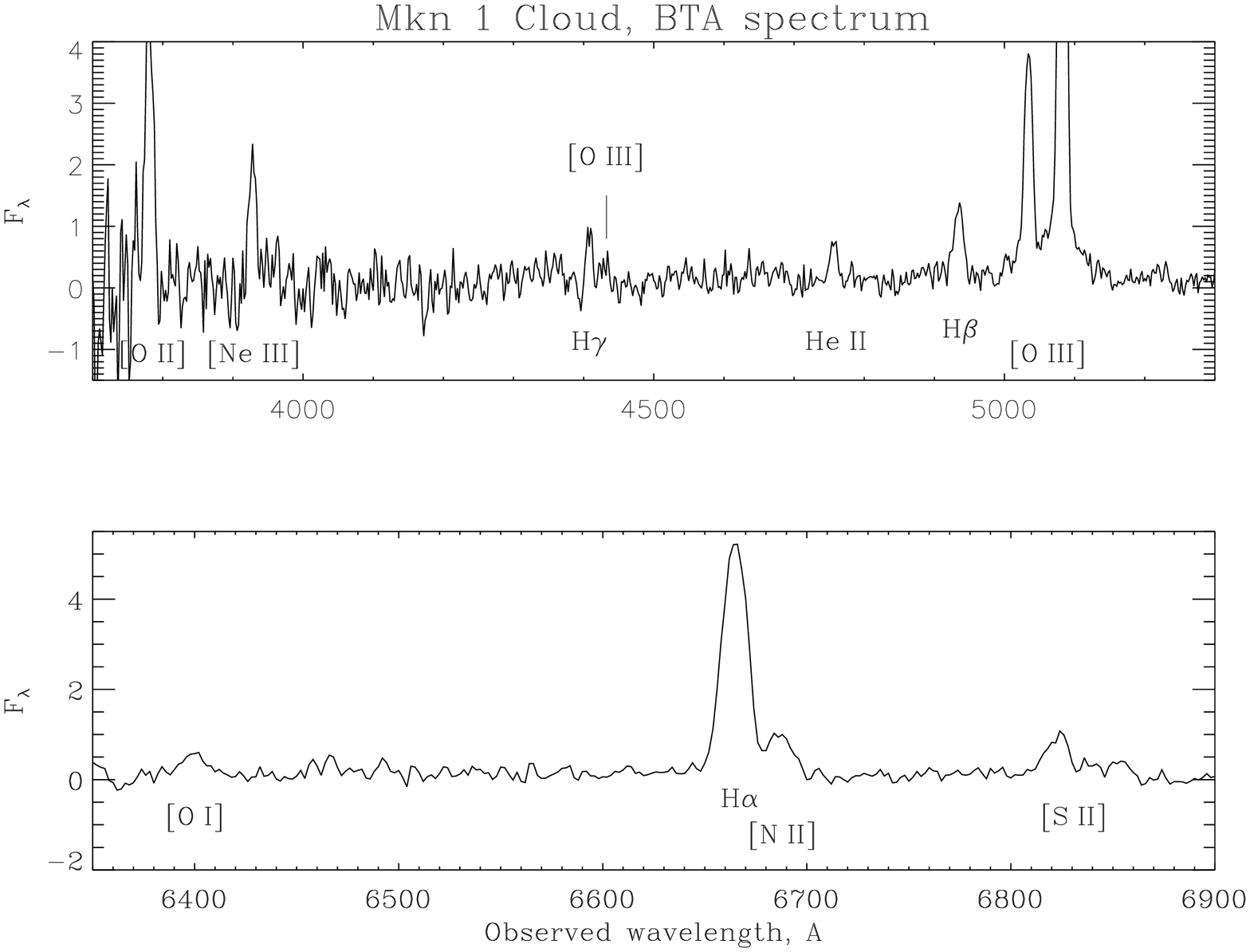} 
\caption{BTA spectrum of the Mkn 1 cloud, summed over a 23\arcsec  region along the 1\arcsec\  slit. The vertical scale,
in units of $10^{-17}$ erg cm$^{-2}$ s$^{-1}$ \AA\ $^{-1}$, 
truncates the peak of [O III] $\lambda 5007$ to show other lines more clearly. That line is 2.93 times as strong as the
$\lambda 4959$ line in energy units \citep{Tayal}.
He II $\lambda 4686$ is clearly detected. The location of the weak, temperature-sensitive [O III] $\lambda 4363$ line is marked. }
 \label{fig-BTAspectrum}
\end{figure*}


\section{Survey results}

Our narrowband images show extended emission-line features interpretable as EELRs in only the case of Mkn 1, as noted above. 

We identified a number of potential compact emission-line sources in the fields of the target galaxies, many in the range of
size and luminosity associated with emission-line dots (ELdots), similar to those seen as outlying star-forming regions
by \cite{RW04} and \cite{Werk2010}.
They are listed in Table \ref{tbl-eldots}, with an indication of whether they are projected inside or outside the lowest-density
H I contours shown by \cite{Kuo}. Detection limits for  emission-line point sources are typically $5 \times 10^{-16}$ erg cm$^{-2}$ s$^{-1}$.

\begin{table*}
\begin{center}
\caption{Compact Emission-Line Candidates}    \label{tbl-eldots}
\begin{tabular}{lccccccc}
\hline\hline
Field-ID & (Possible) Name & z & H I Loc. & $r_{proj}$  & P.A. &  $F(5007)$  \\
   & or Coordinates &   &  &  (arcsec) & ( $\degr$ ) & (erg   cm$^{-2}$ s$^{-1}$) \\
\hline
Mkn 341-II & 2MASX J00364502+2405313 & ... & outside & 389 & 42.8 & $8.7 \times 10^{-15}$\\
Mkn 341-III & SDSS J003706.15+235529.8 & ... & outside & 282 & 138 & $1.1 \times 10^{-14}$\\
Mkn 341-VII & 2MASX J00365081+2355293 & ... & outside & 212 & 185 & $1.2 \times 10^{-14}$\\
Mkn 341-VIII & SDSS J003706.42+235450.5 & ... & outside & 318 & 142 & $1.1 \times 10^{-14}$\\
Mkn 341-X & SDSS J003632.35+240229.4  & ... & outside & 334 & 307 & $5.2 \times 10^{-15}$\\
Mkn 341-XI & SDSS J003631.95+240405.1  & ... & outside & 405 & 317 & $6.4 \times 10^{-15}$\\
Mkn 1-III & near 2XMM J011601.3+330821 & ... & outside & 204 & 345 & $4.1 \times 10^{-14}$\\
NGC 513-I & 2XMM J012437.0+335004 & ... & outside & 134 & 41 & $3.2 \times 10^{-15}$\\
NGC 513-VIII & 2XMM J012434.2+334732 & ... & outside & 105 & 134 & $3.3 \times 10^{-15}$\\
Mkn 1158-I & NCS: 01:35:15.7 +34:56:55 & ... & outside & 376 & 148 & $3.2 \times 10^{-15}$\\
Mkn 1158-II & NCS: 01:34:53.9 +34:57:27 & ... & outside & 306 & 191 & $2.7 \times 10^{-15}$\\
UGC 3157-I & ... & ... & outside & 179 & 246 & $1.4 \times 10^{-14}$\\
UGC 3157-II & knot in arms & ... & inside & 19.7 & 278 & $1.6 \times 10^{-14}$\\
MS 04595+0327-II & NCS: 05:02:16.6 +3:33:46.7 & ... & outside & 160 & 46.0 & $1.1 \times 10^{-13}$\\
MS 04595+0327-III & NCS: 05:01:58.1 +3:29:06.9 & ... & outside & 215 & 223 & $8.5 \times 10^{-14}$\\
MS 04595+0327-IV & NCS: 05:02:01.8 +3:29:15.1 & ... & outside & 187 & 213 & $1.2 \times 10^{-13}$\\
UGC 3995-I & SDSS J074409.03+291228.5 & ... & outside & 320 & 165 & $2.3 \times 10^{-14}$\\
Mkn 1419-I & 145.213571 +3.51826 & ... & outside & 299 & 131 & $4.1 \times 10^{-16}$\\
Mkn 1419-II & SDSS J094107.46+033559.8 & ... & outside & 493 & 81.8 & $4.9 \times 10^{-15}$\\
Mkn 1419-III & SDSS J094053.84+033903.6 & ... & outside & 368 & 44.5 & $8.8 \times 10^{-15}$\\
Mkn 461-III & NCS: 206.83377, +34.18102 & ... & edge & 104 & 14.5 & $1.3 \times 10^{-14}$\\
Mkn 461-IV & NCS: 206.8468, 34.18784 & ... & edge & 186 & 21.0 & $1.3 \times 10^{-14}$\\
NGC 5548-I & SDSS J141818.19+250612.6  & ... & edge(c1) & 281 & 114 & $5.0 \times 10^{-16}$\\
NGC 5548-II & SDSS J141805.94+250416.3 & ... & edge(w) & 258 & 155 & $1.1 \times 10^{-15}$\\
NGC 5548-III & SDSS J141743.50+250400.1 & ... & outside & 320 & 220 & $1.4 \times 10^{-15}$\\
NGC 5548-IV & SDSS J141742.11+250411.4 & ... & outside & 344 & 227 & $1.6 \times 10^{-15}$\\
NGC 5548-VI & 2XMM J141746.9+250725 & ... & outside & 172 & 257 & $1.2 \times 10^{-15}$\\
NGC 5548-VII & SDSS J141745.92+250838.6 & ... & outside & 173 & 275 & $1.5 \times 10^{-15}$\\
NGC 5548-VIII & SDSS J141802.14+251030.4 & ... & outside & 135 & 12.0 & $1.1 \times 10^{-15}$\\
NGC 5548-IX & SDSS J141801.79+251108.3 & ... & outside & 160 & 7.0 & $8.6 \times 10^{-16}$\\
NGC 5548-X & SDSS J141819.78+250913.7 & ... & edge & 348 & 103 & $1.3 \times 10^{-14}$\\
Ark 539-III & 18:29:01.58 +50:23:50.2 & ... & outside & 151 & 54.8 & $4.0 \times 10^{-15}$\\
Ark 539-IV & 18:29:07.51 +50:20:35.4 & ... & inside & 216 & 121 & $4.2 \times 10^{-15}$\\
Ark 539-V & 18:29:07.89 +50:21:33.9 & ... & inside & 194 & 102 & $3.8 \times 10^{-15}$\\
Ark 539-VI & 18:28:57.85 +50:25:37.8 & ... & outside & 223 & 25.3 & $3.5 \times 10^{-15}$\\
\hline
\end{tabular}
\end{center}
\end{table*}

\section{Mkn 1 and its distant ionized cloud}

\subsection{Ionization}
Our initial Lick spectrum showed the cloud to have high ionization, but left the ionizing mechanism ambiguous. The deeper BTA
data show He II emission at a level requiring an AGN continuum if photoionized, and also show that the electron temperature and line widths are low enough to strongly disfavor shock ionization. This object is an instance of low-metallicity gas which can masquerade as
being ionized by hot stars in the single strong-line BPT \citep{BPT} diagram 
\citep{Groves2006}.
For this reason, detection
of He II emission has been important in establishing the ionizing source of extended clouds seen near galaxies, showing in some cases
the presence of either obscured or fading AGN (\citealt{Lintott2009}, \citealt{mnras2012}, \citealt{ngc7252}, \citealt{Keel2019}).

Table \ref{tbl-lineratios} shows emission-line ratios integrated along the slit for the entire cloud, and for the two
subregions seen in the direct image. Modest differences in ionization properties are found between the NW and SE regions,
with the SW part of the cloud more highly ionized, as seen in [O III]/H$\beta$ and He II/H$\beta$. Good agreement is seen 
between integrated line ratios from our two sets of spectral data.

The [O III] $\lambda 4363$ line is marginally detected (formally at the 2$\sigma$ level) in the cloud integrated spectrum
(Fig. \ref{fig-BTAspectrum}). This lets us place a lower limit on the
electron temperature via its ratio with the $\lambda 5007$ line. Using the {\it temden} task in the {\it nebular} package within
IRAF \citep{ShawDufour}, we place a $3 \sigma$ bound $T_e > 20,000$ K for low electron density ($n_e<50$ as is typical for such distant AGN clouds). This temperature, and the ionization level found from [O III]/H$\beta$ and He II/H$\beta$, clearly point to AGN radiation as the
ionizing agent for this cloud. The line ratios are very similar to those observed in the EELRs studied in the Galaxy Zoo sample, as
listed in Table 4 of \cite{mnras2012} with the mean values shown in Table \ref{tbl-lineratios} here.

\begin{table*}
\begin{center}
\caption{Spectroscopic Emission-line Ratios}  \label{tbl-lineratios} 
\begin{tabular}{lcccccc}
\hline
Ratio & Lick& BTA Integrated & SE & NW& Mean EELR \\\\
\hline
\ [O II] $\lambda 3727$/[O III] $\lambda 5007$ & 0.42 & 0.39 & 0.43 & 0.50 & 0.51 \\
\ [O III]$\lambda 5007$/H$\beta$   &  -- & 9.34  & 10.9 &  7.20& 8.38& \\
\ [O III]$\lambda 5007$/H$\alpha$ & 2.40 & 2.22 & 2.83 & 1.47 & 2.74 \\
\ H$\alpha$/H$\beta$ &   --    &     4.20  &  3.85 & 4.90& --\\
\ [N II]$\lambda 6583$/H$\alpha$&    0.185 &         0.16  & 0.20 &  0.127 & 0.43\\
\ [O I]$\lambda 6300$/H$\alpha$ &   --  &        0.085 & 0.093 &  0.076& 0.071\\
\ He II$\lambda 4686$/H$\beta$      & --   &    $0.43\pm 0.06$ &   $0.61\pm 0.12$ &   $0.28\pm 0.12$ & 0.30 \\
\ [Ne III] $\lambda 3869$/ O II] $\lambda 3727$ & -- & 0.39 &   0.40  & 0.48 & 0.33\\
\hline
\end{tabular}
\end{center}
\end{table*}


\subsection{Relation to circumnuclear gas and H I }
\cite{Stoklasova} present integral-field spectroscopy of the central regions of Mkn 1, showing resolved emission lines across
a region spanning about 20\arcsec 
along the projected major axis of the inner disc of Mkn 1. Using the [O III]/H$\beta$ ratio to identify potential ionization cones shows narrow high-ionization regions to
NE and SW. Outside the nuclear emission extending to about $r=4$\arcsec\ , the possible ionization cones are seen between position angles
50--85$^\circ$ with the highest ionization along PA 65$^\circ$, slightly misaligned with the stellar distribution.

Archival {\it Hubble Space Telescope} images \citep{Pjanka} show the inner disc to be elongated along PA 72$^\circ$, with numerous star-forming
regions most prominent on its south and western sides (contributing strongly to the Balmer H$\alpha$ and H$\beta$ flux maps from \citealt{Stoklasova}).

The distant cloud we have found is projected at radii 33--45\arcsec\  and spans PA 207--236$^\circ$, while the inner possible ionization cone
on this side spans 230--265$^\circ$, barely overlapping. If both structures are ionized by radiation in a fixed conical pattern, neither
structure samples most of the cone width. This might mean that the inner disk gas and the outer H I structure are misaligned
with each other and thus intercept different portions of the ionizing radiation pattern. Alternately,
precession of the ionizing pattern on time scales $\approx 3 \times 10^4$ years could produce this offset.

The highest-resolution available H I data are the GMRT results from \cite{Omar}, with beam size $27 \times 31$\arcsec FWHM. 
Fig. \ref{fig-OmarOverlay} schematically
shows the locations of the southwest and possible northeast emission clouds compared to their H I contours. The confirmed SW 
cloud falls right at the outer edge of the H I structure, possible associated with an inflection of declining surface density,
while the possible NE cloud falls largely in the gap between the two H I features in that side. We might speculate that
this is similar to the location of Hanny's Voorwerp in a gap in the H I tail of IC 2497, where ionization of most of the
gas creates a gap in the H I structure. Fig. \ref{fig-OmarOverlay} also illustrates how the outer clouds and inner ionization
cones overlap only partially in orientation around the nucleus.

\begin{figure*} 
\includegraphics[width=120.mm,angle=90]{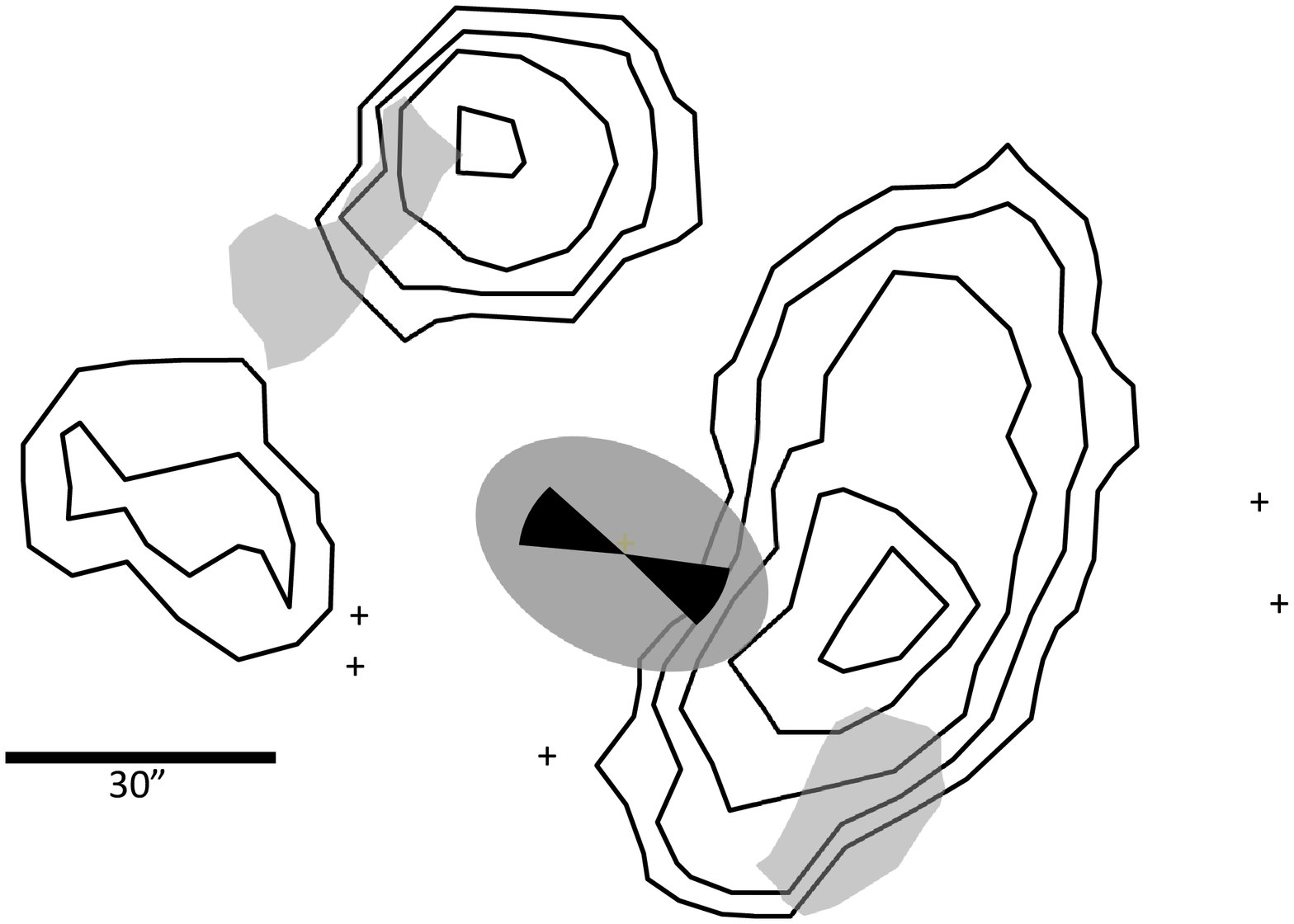} 
\caption{Locations 
of the [O III] emission clouds , shown shaded within the lowest contours from
from Fig. \ref{fig-Mkn1Byu}, and H I emission contours taken graphically from Fig. 5 of 
Omar et al.,
aligned using the star positions shown by crosses. Contour levels are $(0.3, 0.8, 1.3, 2.3, 3) \times 10^{20}$ cm$^{-2}$.
The ellipse shows the optical disk of Mkn 1, within much of which H I emission will be hidden by 
absorption against the AGN continuum source. North is at the top, east to the left; the scale bar indicates the
major axis of the synthesized H I beam, which has axis ratio 0.89 elongated in position angle $100^\circ$.
The black zones within the schematic disk show the extent of the ionization cones stupid by Stoklasova et al., and
their partial misalignment with the more distant clouds we have identified.}
 \label{fig-OmarOverlay}
\end{figure*} 

\subsection{Related dwarf galaxy?}
A diffuse continuum object appears projected near the edge of the northwestern component of the emission-line cloud. As revealed
by the HST WF3 F814W image \citep{Pjanka}, there is only one other similar bright and diffuse object in the Mkn 1 field, a nucleated
object of comparably low surface brightness on the opposite side of Mkn 1 itself. This raises at least the possibility that
this is a dwarf companion to Mkn 1, and the AGN is ionizing gas associated with this companion rather than
purely diffuse H I in its surroundings (which would be an instance of cross-ionization, in the terminology of \citealt{Keel2019}).
The relation between the emission cloud and continuum objects is shown in Fig. \ref{fig-SARAonHST}. The ionized gas
component is  not centered on the galaxy, leaving the possible association ambiguous.

\begin{figure*} 
\includegraphics[width=120.mm,angle=270]{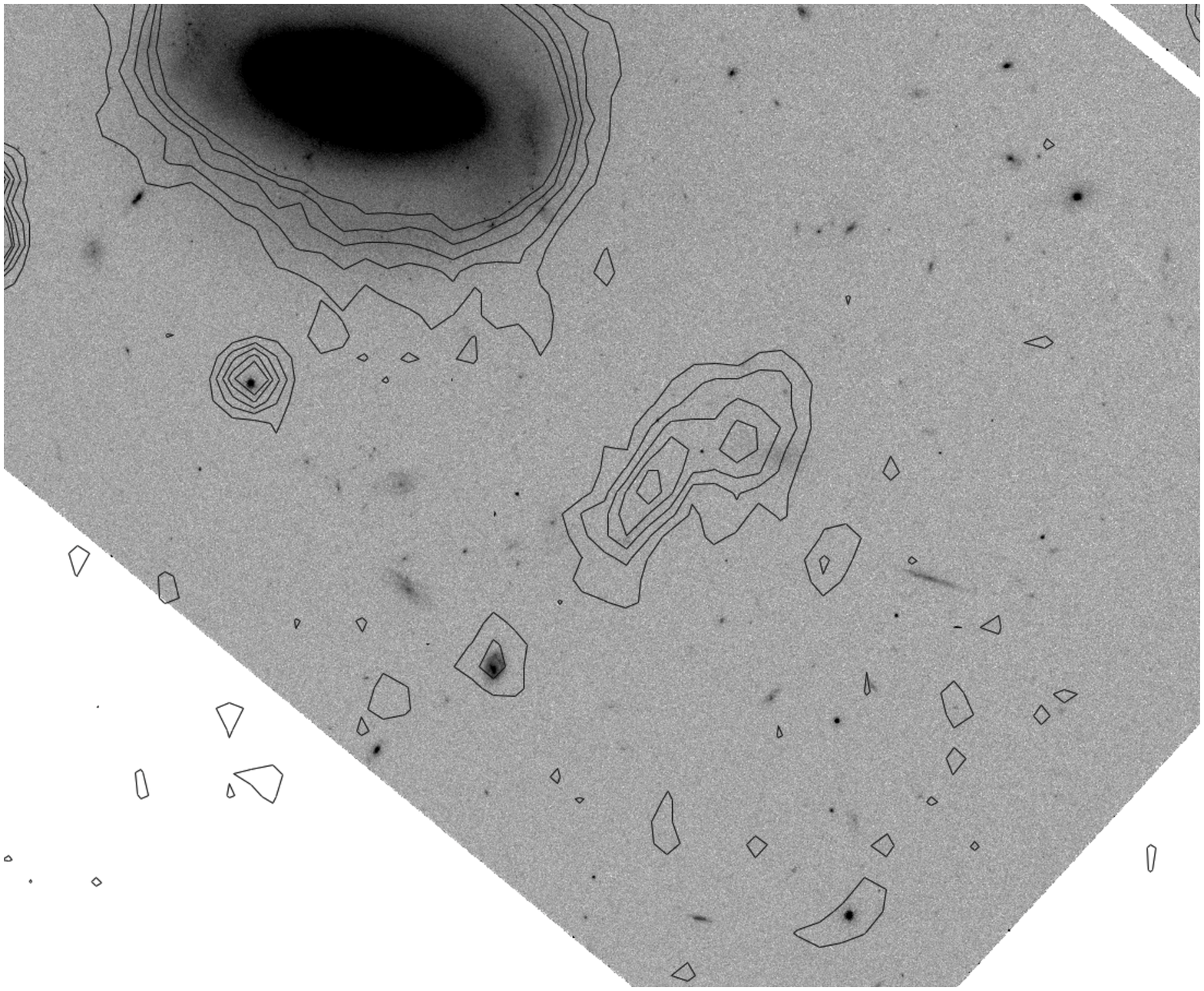} 
\caption{Contours of the narrowband SARA [O III] image of Mkn1 overlaid on
archival HST WCF3 F814W image, showing the relation between the 
emission-line cloud and diffuse continuum object. Coordinates for the narrowband image were refined
by about 1\arcsec\  to match locations of stars in the HST image.
The region shown is $76 \times 94$ \arcsec with north at the top.}
 \label{fig-SARAonHST}
\end{figure*} 

\subsection{Kinematics}
We examine the velocity structure of the cloud by using wavelength peaks as retrieved from Gaussian fitting of the strongest
lines, constraining both [O III] lines to have the same width. The results show a gradient of about 60 km s$^{-1}$ along
the detected extent of the cloud, with both the radial velocity and gradient in good agreement with the H I as measured by
\cite{Kuo} and \cite{Omar}. The velocity gradient in both H I data set is shallower, but subject to beam smoothing.. The gradient is subtle enough, in view of our typical uncertainty of order 30 km s$^{-1}$, that it is unclear whether
what two apparently distinct structural components in the cloud are also kinematically distinct (Fig. \ref{fig-Mkn1velslice}).
The velocity data weakly suggest that the NW component has constant velocity while the SE one shows a gradient, but a
single linear gradient is almost as good a fit. 

\begin{figure} 
\includegraphics[width=65.mm,angle=90]{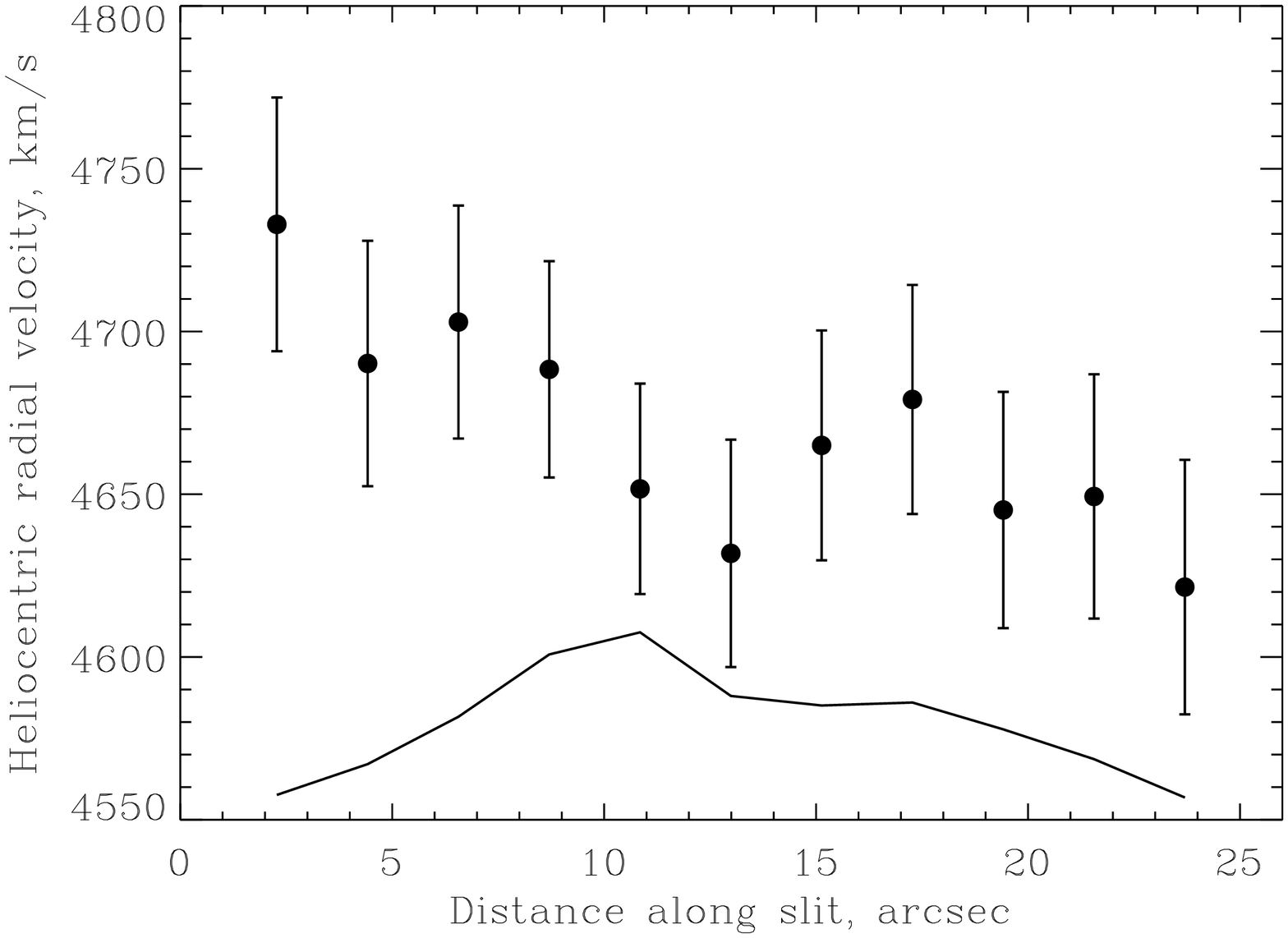} 
\caption{Radial-velocity structure of the Mkn 1 cloud along the BTA spectroscopic slit, evaluated by Gaussian fits
summed over 2\arcsec\  regions. The continuous curve at the bottom shows the intensity of [O III] emission, illustrating how
the two structural components may have different velocity gradients. Distance along the slit increases from southeast to northwest.}
 \label{fig-Mkn1velslice}
\end{figure} 

\subsection{Energy budget and AGN history}

In principle, the ionization level of outlying EELR gas samples the past luminosity of the AGN. 
We follow the approach of \cite{mnras2012} to estimate the minimum AGN luminosity needed to power the distant cloud, and compare with the obscured AGN luminosity (since Mkn 1 is a Type 2 Seyfert, this is appropriate). The cloud's H$\beta$ surface rightness and projected distance from 
the nucleus require an ionizing luminosity $ > 10^{43}$ ergs s$^{-1}$, by a factor depending on the clumpiness of the gas (and through this, ultimately
connected to its optical depth at the Lyman limit). The upper limit to the electron density from the [S II] lines $n_e < 26$ cm$^{-3}$ can give
a complementary upper limit to the ionizing luminosity. Following \cite{KomossaSchulz}, the ionization parameter $U$ derived from the
[O II]/[O III] line ratio (Table \ref{tbl-lineratios}) and a typical AGN continuum shape is $U=0.0032$. The emission rate of ionizing
photons is $Q_{ion)} = U/(4 \pi r^2 n_e)$, so for $r=13$ kpc and $n_e<26$, $Q_{ion} < 5.0 \times 10^{53}$ photons s$^{-1}$.
For a generic AGN continuum, the mean energy of ionizing photons is roughly 2 Rydbergs, so this becomes 
$L_{ion} < 2.2 \times 10^{43}$ erg s$^{-1}$. These two approaches give bounds on the ionizing luminosity seen by the cloud at the epoch 
of the emitted photons $1.6 \pm 0.6 \times 10^{43}$ erg s$^{-1}$.

The far-infrared luminosity, predominantly from reprocessed UV radiation, is 
$1.7 \times 10^{44}$ ergs s$^{-1}$. A more complete accounting, adding the escaping ionizing radiation as traced by [O III]
emission from the AGN itself,
gives $2.1 \times 10^{44}$ ergs s$^{-1}$. This is sufficiently greater than the ionizing requirements for the cloud to make it plausible 
(although not required) that the AGN 
has maintained a roughly constant output over the light-travel time to the clouds ($\approx 40,000$ years).

\section{Sample implications for AGN cone angles and episode durations}

We give a simple calculation of the typical width of ionization cones, under some simplifying assumptions. These are
(1) that all the AGN are powerful enough to ionize observable clouds in the H I structures, and (2) that these structures are
reasonably approximated as great-circle arcs viewed from each AGN. Estimates of the ionizing luminosity of each AGN,
constructed from the far-IR luminosities and nuclear [O III] luminosities, lie with a factor $\approx 3$ of the value for Mkn 1 except for 
Akn 539, Mkn 461, UGC 3157, NGC 1167, and NGC 841 at lower luminosity.
We used the published H I column-density maps from \cite{Kuo} to estimate the angle  $A$ spanned by H I features around each AGN, 
approximated as tangential arcs (Table \ref{tbl-HI}, along with the estimated AGN ionizing luminosity). Details of calculating the chances of 
randomly distributed arcs and (bi)cones intersecting are given in the Appendix. 

Most of the nearby H I structures are associated with companion galaxies, and the H I data give redshifts making physical
association very likely (so this is a search for cross-ionization as defined by \citealt{Keel2019}).
Projection factors in these cases are poorly constrained, with no necessary connection to the AGN host inclination. In view of this,
we considered examination of projection effects based on the inclinations of the remaining hosts, but as this would affect only 15\%
of the entries in Table \ref{tbl-HI}, the outcome would still be dominated by the behavior of gas around companion galaxies.

\begin{table}
\begin{center}
\caption{Extended H I Structures}    \label{tbl-HI}
\begin{tabular}{l l l c l}
\hline\hline
Field &  $r_{proj}$  &  A &  $L_{ion}$   &  $ A_{half}$ \\ 
   & (kpc) & (\degr) &  (erg  s$^{-1}$) &  (\degr) \\
\hline
NGC 7469 & 24.7 & 120 & $3.0 \times 10^{44}$ & 31.9  \\ 
NGC 5548-1 & 54.5 & 23 & $2.5 \times 10^{44}$ & 53.0  \\
NGC 5548-2 & 80.5 & 15 & $2.5 \times 10^{44}$ & 55.3  \\ 
NGC 5548-3 & 20.9 & 79 & $2.5 \times 10^{44}$  & 39.1  \\ 
Ark 539 & 54.1 & 77 & $4.7 \times 10^{42} $ & 39.5  \\ 
NGC 7679 & 62.3 & 72 & $1.1 \times 10^{44}$ & 40.5  \\ 
NGC 7682 & 62.3 & 52 & $2.4 \times 10^{44}$ & 45.1  \\ 
Mkn 461 & 58.9 & 26 & $2.6 \times 10^{43}$ & 52.0  \\
MS 04595+0327 & 47.4 & 36 & $7.2 \times 10^{44}$  & 49.4  \\ 
UGC 3157 & 68.5 & 26 & $9.4 \times 10^{42}$ & 52.0  \\ 
Mkn 993 & 82.2 & 16 & $9.0 \times 10^{42}$  & 55.2  \\ 
NGC 513 & 47.4 & 57 & $5.8 \times 10^{43}$  & 44.0  \\ 
Mkn 1 & 62.3 & 74 & $2.1 \times 10^{44}$ &   40.1  \\
NGC 1167 & 82.2 & 8.2 & $2.3 \times 10^{42}$ &   57.5  \\
UGC 1395 & 52.1 & 31 & $2.5 \times 10^{44}$ &   50.7  \\
Mkn 573 & 75.4 & 10 & $4.7 \times 10^{44}$ &   56.9  \\
Mkn 1157 & 41.9 & 20 & $5.8 \times 10^{43}$ &   53.9  \\
NGC 7591 & 50.0 & 26 & $6.3 \times 10^{43}$ &   52.0  \\
NGC 841 & 109 & 24 & $7.1 \times 10^{41}$ &   52.8  \\
Mkn 341  & 76.0 & 12 & $3.0 \times 10^{43}$ &   56.2  \\
\hline
\end{tabular}

Notes: {$A_{half}$ is the angle where $P(A,L)=\frac{1}{2}$ as derived in the Appendix; with the observed angle $A$ (col. 3) measured on H I map} 
\end{center}
\end{table}

For the entire sample, the number of detections for the observed arc angles would suggest narrow cones of escaping
ionizing radiation, typically $20^\circ$ width. This is so much narrower than most observed cone angles that it may
suggest intermittent luminous episodes.

This conceptually clean formulation runs into several complications in actual data, so our results also provide
an outline of ways to improve this approach. Projection effects can increase or decrease measured arc lengths,
and could render an arc of gas more like a radial spur or blend it with the host galaxy disc. The limited angular resolution of
the current H I data often blend external structures, to some extent, with the host disc gas. Finally, shadowing of 
some of the H I by material in the inner disc or AGN torus will often prevent ionizing radiation from escaping in both the planes
(broadened if either one is warped or twisted; \citealt{LawrenceElvis}). Our primary interest is in ionized gas outside the
normal disc interstellar medium of the host galaxy; a sufficiently warped disc could escape the self-shielding of the inner regions. 
Lacking a prescription for a fuller treatment, we present the results as is with these cautionary notes.

\section{Conclusions}

We have surveyed the environments of a set of Seyfert galaxies with known H I structures, using a narrowband [O III] filter,
in search of distant nebulosities photoionized by the AGN (EELRs). Among the 26 Seyfert galaxies with H I observations by
\cite{Kuo}, we find an EELR in one case. Mkn 1 is accompanied by a large emission-line region projected 12 pc from the
AGN, for which spectroscopy confirms that it is photoionized by the AGN itself. Radial velocities support an association between
this material and the larger extended H I disc. The location of this ionized region does not match the ionization cone inferred from
circumnuclear gas, suggesting either that neither emission region fully crosses the escaping cone of radiation or that
the cone of radiation moves on timescales $\approx 30,000$ years. A fainter emission-line cloud may appear on the opposite side of the
galaxy, but interference from a very bright foreground star leaves this detection tentative.

Finding one EELR out of 20 extended H I structures in 18 galaxies suggests either
that typical ionization cones are quite narrow (opening angle $\approx 20 ^\circ$) or that 
bright AGN phases are episodic compared to light-travel times 20,000-50,000 years. If the AGN in our sample are luminous enough to ionize
gas at these distances, as seen from the example of Mkn 1 and nuclear luminosities in Tale \ref{tbl-HI}, this conclusion follows from the expected number of cases in which
randomly directed bicones intercept gaseous arcs of the lengths inferred from the H I data.

\section*{Acknowledgements}

The observations at the SAO RAS 6 m telescope were carried out with the financial support support of the Ministry of Science and Higher Education of the Russian Federation
VNB acknowledges assistance from National Science Foundation (NSF)
Research at Undergraduate Institutions (RUI) grant AST-1312296. GK acknowledges support from
NSF grants DMS-1363239 and DMS-1900816. Findings and conclusions
do not necessarily represent views of the NSF. AM, SD  and  AG thank the grant of Russian Science Foundation project 17-12-01335 
``Ionized gas in galaxy discs and beyond the optical radius".

Funding for the creation and distribution of the SDSS Archive has been
provided by the Alfred P. Sloan Foundation, the Participating Institutions,
the National Aeronautics and Space Administration, the National Science
Foundation, the U.S. Department of Energy, the Japanese Monbukagakusho,
and the Max Planck Society. The SDSS Web site is http://www.sdss.org/. 
The SDSS is managed by the Astrophysical Research Consortium (ARC) for
the Participating Institutions. The Participating Institutions are The
University of Chicago, Fermilab, the Institute for Advanced Study, the Japan
Participation Group, The Johns Hopkins University, Los Alamos National
Laboratory, the Max-Planck-Institute for Astronomy (MPIA), the
Max-Planck-Institute for Astrophysics (MPA), New Mexico State
University, Princeton University, the United States Naval Observatory, and
the University of Washington.

This research has made use of the NASA/IPAC Extragalactic Database (NED),
which is operated by the Jet Propulsion Laboratory, Caltech, under
contract with the National Aeronautics and Space Administration. IRAF is distributed by the National Optical Astronomy Observatory, which is operated by the Association of Universities for Research in Astronomy (AURA) under a cooperative agreement with the National Science Foundation. The authors are honored to be permitted to 
conduct astronomical research on Iolkam Du'ag (Kitt Peak), a mountain with particular significance to the 
Tohono O'odham Nation.

\appendix

\section{Probability of an arc intersecting a double cap}

Given a (double) cone of opening angle $A \leq \pi/2$ intersecting the unit sphere $x^2+y^2+z^2=1$ at two caps, we wish to find the
probability $P(A,L)$ that a random geodesic arc of length $L\leq 2\pi$ intersects one or both of the caps.  In spherical coordinates, the
caps are the regions $0\leq \phi \leq A$ and $\pi - A \leq \phi \leq \pi$.  Here we use spherical coordinates $\theta, \phi$ so that $0\leq
\theta \leq 2\pi$, $0\leq \phi \leq \pi$ and $x = \cos \theta \sin \phi$, $y=\sin\theta \sin \phi$, $z = \cos \phi$.

We shall compute the probability using conditional probabilities.  Let $C$ denote the center of the random geodesic arc and $\theta(C),
\phi(C)$ the corresponding spherical coordinates.  Let $P(A,L,\phi)$ denote the probability that the arc intersects a cap given that
$\phi(C) = \phi$.  Then, \begin{equation} 
\label{pal} 
P(A,L) = \int_{0}^{\pi} P(A,L,\phi) \frac{1}{2}\sin \phi d\phi =
2\int_{0}^{\pi/2} P(A,L,\phi) \frac{1}{2}\sin \phi d\phi 
\end{equation}
where $\frac{1}{2} \sin \phi d\phi$ is just the probability density
for $\phi(C)$.  Namely, the probability that $a<\phi(C)<b$ is obtained
by integrating $\frac{1}{4\pi} \int_{a}^{b}\int_{0}^{2\pi} \sin \phi
d\theta d\phi = \int_{a}^{b} \frac{1}{2} \sin\phi d\phi$ which is the
proportion of the sphere in the range $a<\phi < b$.

The expression on the right in Eqn. \ref{pal} reflects the symmetry that
the probabilities are the same for $\phi$ and $\pi-\phi$.  So, we only
need to consider $0\leq \phi \leq \pi/2.$

Some simple cases are recorded below
\[
P(A,L,\phi) = \begin{cases} 0 & \text{ if } A+L/2 \leq \phi \leq \pi/2 \\
1 & \text{ if } 0\leq \phi \leq A  \\
? & \text{ if } A < \phi < \min\{\pi/2, A+L/2\} \\
\end{cases}
\]
The first case corresponds to values of $\phi$ where the arc cannot
reach a cap no matter which way it points, the second case corresponds
to when the center of the arc lands in a cap.  The third case will now
be determined.

To begin, there is no harm in assuming $\theta(C) = 0$ since the
problem is rotationally symmetric.  Then, $C = (\sin \phi, 0 , \cos
\phi)$.  The vectors $\vec{a} = \langle \cos \phi, 0, - \sin
\phi\rangle$, $\vec{b} = \langle 0 ,1,0\rangle$, $\vec{c} = \langle
\sin \phi, 0 ,\cos \phi\rangle$ are pairwise perpendicular and can be
used to form a local coordinate system at $C$ (they are a rotated
version of $\mathbf{i},\mathbf{j},\mathbf{k}$).  The arc centered at
$C$ can be parametrized via
\[
\vec{c}(t) = (\sin t \cos \theta) \vec{a} + (\sin t \sin \theta)
\vec{b} + (\cos t) \vec{c}.
\]
This is derived by treating $\vec{a}, \vec{b},\vec{c}$ as we would
usually treat $\mathbf{i},\mathbf{j},\mathbf{k}$ in spherical
coordinates.  Here $-L/2 \leq t \leq L/2$ and $\theta$ will now be
used to represent the angle the arc makes with $\vec{a}$.  By symmetry
we may assume $0 \leq \theta \leq \pi/2$.

It helps to recall the inequality 
\begin{equation} \label{ineq}
|a \sin t + b \cos t| \leq \sqrt{a^2+b^2}.
\end{equation}
Equality occurs exactly when 
\begin{equation} \label{equal}
(\sin t, \cos t) =
\frac{\pm 1}{\sqrt{a^2+b^2}} (a,b).  
\end{equation}

The great circle containing our geodesic arc intersects a cap if there
is a value of $t$ such that the $z$-coordinate of $\vec{c}(t)$ is $\pm
\cos A$; i.e.
\begin{equation} \label{zcoord}
\pm \cos A = -\sin t \cos \theta \sin \phi + \cos t \cos \phi.
\end{equation}
Setting $a=-\cos \theta \sin\phi$ and $b = \cos\phi$ in \eqref{ineq},
equation \eqref{zcoord} has a solution exactly when $\cos^2 A \leq
\cos^2 \theta \sin^2 \phi + \cos^2 \phi = 1- \sin^2 \theta \sin^2
\phi$.  This is equivalent to $\sin \phi \sin \theta \leq \sin A$, and
the maximal possible angle $\theta$ such that the great circle
intersects a cap is $\theta_{max} = \sin^{-1}(\frac{\sin A}{\sin \phi})$.
The corresponding values of $t$ satisfy
\begin{equation} \label{sincos}
(\sin t, \cos t) = \frac{\pm
  1}{\cos A}(-\cos (\theta_{max}) \sin \phi, \cos \phi) \text{ by \eqref{equal}.}
\end{equation}
 
 Since $A,\phi,\theta_{max}$ are all in the range $[0,\pi/2]$, $\sin t$
  and $\cos t$ have opposite signs which means $(\sin t, \cos t)$ is
  in quadrants IV and II.  Therefore, the closest point of
  intersection occurs when $-\pi/2 \leq t \leq 0$ and this means the
  great circle intersects both caps or no caps.  In particular, if
  $L/2 \geq \pi/2$, then the arc intersects the cap for $\theta$ in
  the range $0\leq \theta \leq \sin^{-1}(\frac{\sin A}{\sin\phi})$.

This shows that for $L\geq \pi$
\[
P(A,L) 
= 2(\frac{1}{\pi} \int_{A}^{\pi/2} \sin^{-1}(\frac{\sin A}{\sin \phi})
\sin \phi d\phi + \int_{0}^{A} \frac{1}{2}\sin{\phi} d\phi)
\]
which does not depend on $L$.

For $L < \pi$, the arc intersects the cap in the full sweep $0\leq
\theta \leq \theta_{max}$ as long as the intersection point at the maximal
angle occurs for $|t|\leq L/2$, or equivalently $\cos t \geq \cos
L/2$.  By \eqref{sincos}, $\cos t = \frac{\cos\phi}{\cos A}$ at the
maximal angle, and when $\cos L/2 \leq \frac{\cos\phi}{\cos A} = \cos
t$ the arc intersects the cap for $0\leq \theta \leq
\sin^{-1}\frac{\sin A}{\sin \phi}$. Namely, for each $\phi$ with $A
\leq \phi \leq \cos^{-1}(\cos A \cos(L/2))$, we get $P(A,L,\phi) = \frac{2}{\pi}
\sin^{-1} \left(\frac{\sin A}{\sin \phi}\right)$.

If the arc does not intersect the cap in the full range $0\leq \theta
\leq \theta_{max}$, then $\phi > \cos^{-1}(\cos A \cos(L/2))$.  Assuming
also $A+L/2 \geq \phi$, so that the arc can actually intersect the
cap, the maximal angle at which the arc intersects the top cap occurs
when $t = -L/2$.  From \eqref{zcoord}, when $t=-L/2$
\[
\cos \theta = \frac{\cos A - \cos(L/2)\cos \phi}{\sin (L/2) \sin \phi}
\]
making the range of $\theta$ given by $0\leq \theta \leq
\cos^{-1}(\frac{\cos A - \cos(L/2)\cos \phi}{\sin (L/2) \sin
  \phi})$. So, $P(A,L,\phi) = \frac{2}{\pi} \cos^{-1}(\frac{\cos A -
  \cos(L/2)\cos \phi}{\sin (L/2) \sin \phi})$ for $\cos^{-1}(\cos A
\cos(L/2)) < \phi \leq \min\{\pi/2, A+L/2\}$.

We can now compute $P(A,L)$ according to two cases.

For $L \geq \pi$
\[
P(A,L) 
= \frac{2}{\pi} \int_{A}^{\pi/2} \sin^{-1}(\frac{\sin A}{\sin \phi})
\sin \phi\ d\phi + 1-\cos A.
\]
For $L< \pi$ 
\[
\begin{aligned}
P(A,L)=& 
1- \cos A + \frac{2}{\pi}\int_{A}^{\cos^{-1}(\cos L/2 \cos A)}
\sin^{-1}(\frac{\sin A}{\sin \phi}) \sin \phi \ d\phi \\
&+ \frac{2}{\pi}
\int_{\cos^{-1}(\cos (L/2)\cos A)}^{\min\{\pi/2, A+L/2\}} \cos^{-1}\left( \frac{\cos
  A - \cos (L/2) \cos \phi}{\sin (L/2) \sin \phi}\right) \sin \phi\ 
d\phi
\end{aligned}
\]

\begin{figure} 
\includegraphics[width=65.mm,angle=90]{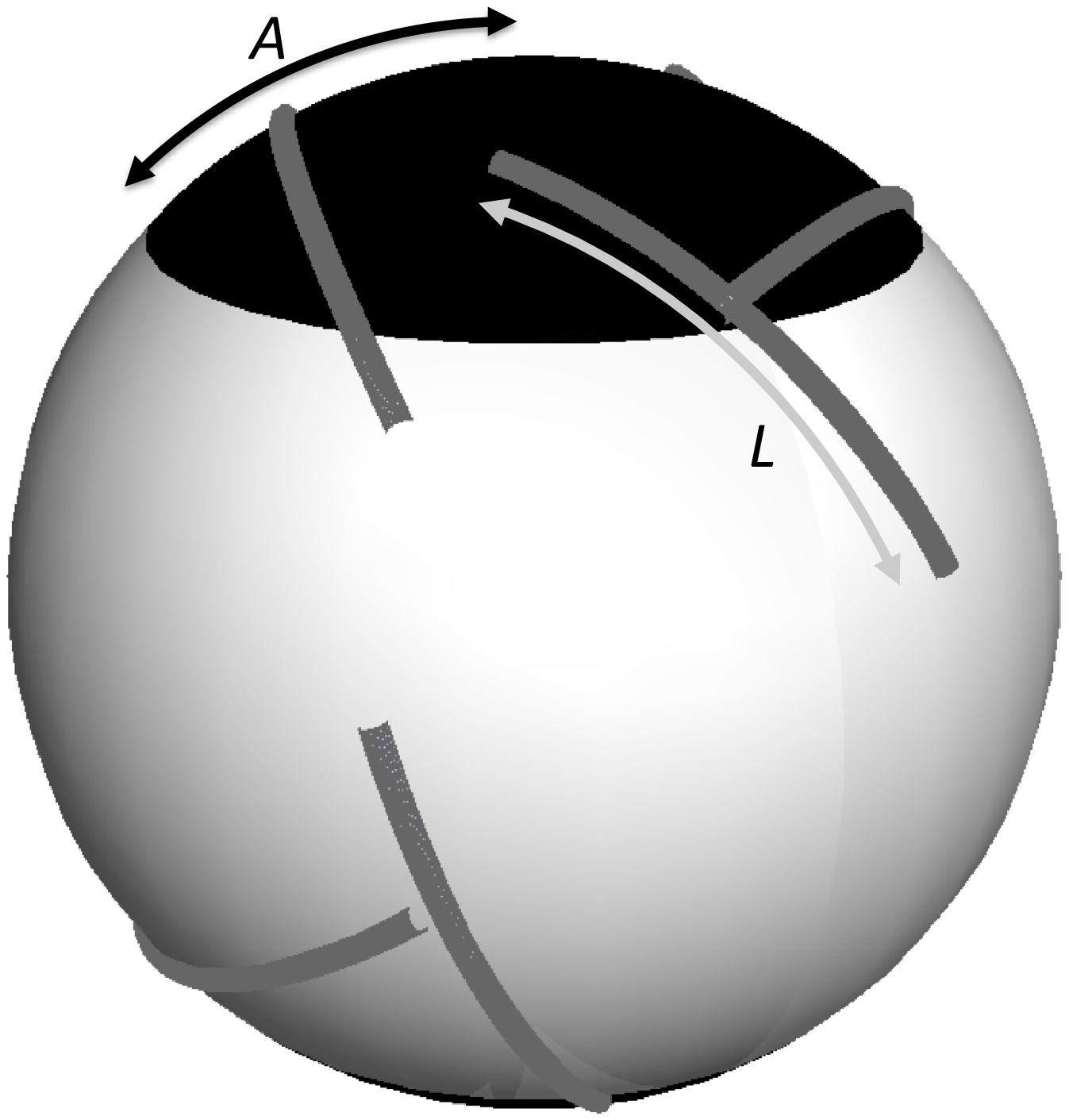} 
\caption{Illustration of quantities used in the bicone-arc intercept calculation. As projected on a unit sphere, the black shading shows the cap 
spanned by one half of the notional ionization (bi)cone, with opening angle $A$. The dark gray arcs are randomly oriented with angular 
length $L$, representing in our study the extent of H I outside the host galaxy as seen by the nucleus. We aim to determine the
typical value of $A$ based on using ionized gas to trace the number of intercepts of arcs of various measured angular extent, based on
the assumption that ionized gas will reveal etches of such intercepts.
}
 \label{fig-appendix}
\end{figure}

\label{lastpage}

\end{document}